\documentclass[
amsmath,amssymb,
aps,
prd, 
twocolumn, 
amsmath,  
nofootinbib,
superscriptaddress,
showkeys
]{revtex4-2}
\usepackage[utf8]{inputenc}
\usepackage{graphicx}% Include figure files
\usepackage[T1]{fontenc}
\usepackage{float}
\usepackage{color}
\usepackage{amsmath}
\usepackage{multirow}
\usepackage{hyperref}% add hypertext capabilities
\makeatletter\def\Hy@Warning#1{}\makeatother
\usepackage[normalem]{ulem}
\hypersetup{colorlinks=true, citecolor=blue}

\newcommand{\chav}[1]{\ensuremath{\left\{#1\right\}}}

\begin{document}

\title{Constraining neutron star matter from the slope of the mass-radius curves}

\author{Márcio Ferreira}
\email{marcio.ferreira@uc.pt}
\affiliation{CFisUC, 
	Department of Physics, University of Coimbra, P-3004 - 516  Coimbra, Portugal}
	
\author{Constança Providência}
\email{cp@uc.pt}
\affiliation{CFisUC, 
	Department of Physics, University of Coimbra, P-3004 - 516  Coimbra, Portugal}

\date{\today}

\begin{abstract}
We analyse the implications of information about local derivatives from the mass-radius diagram in neutron star matter. It is expected that the next generation of gravitational wave and electromagnetic detectors will allow the determination of the neutron star radius and mass with a small uncertainty.  Observations of neutron stars clustered around a given neutron star mass allow the estimation of local derivatives in the $M(R)$ diagram, which can be used to constrain neutron star properties. From a model-independent description of the neutron star equation of state, it is shown that a $M(R)$ curve with a negative slope at 1.4$M_\odot$ predicts a $2M_\odot$ neutron star radius below 12 km. Furthermore, a maximum mass below 2.3$M_\odot$ is obtained if the $M(R)$ slope is negative in the whole range of masses above $1M_\odot$, and a maximum mass above 2.4$M_\odot$ requires the $M(R)$ slope to be positive in some range of masses.   Constraints on the mass-radius curve of neutron stars will place strong constraints on microscopic models.\end{abstract}

\maketitle

%\tableofcontents

\section{\label{sec:introduction} Introduction}

Many studies have been carried out to determine the high-density equation of state (EoS) from the knowledge of the mass-radius curve of the neutron star (NS) \cite{Raithel:2017ity,Fujimoto:2019hxv,Lim:2020zvx,Raaijmakers:2021uju,Huang:2023grj,Carvalho:2024kgf}.  The determination of the mass and radius of an NS with sufficient precision will allow the high-density baryon EoS to be constrained. The Neutron Star Interior Composition
Explorer (NICER) is expected to measure the mass and radius of an NS with an uncertainty of 5\% \cite{Watts:2024ozl}. Future detectors such as the Enhanced X-ray Timing and Polarimetry mission (eXTP) \cite{eXTP,eXTP:2018anb}, the (STROBE-X) \cite{STROBE-X} will improve this precision to $\sim$2\%. Constraints on NS mass and radius are also expected from the detection of gravitational waves emitted by binary NS mergers, such as GW170817 \cite{LIGOScientific:2018cki} reported by the LIGO/Virgo collaboration. The third generation of gravitational wave detectors is expected to allow the determination of the NS radius with an uncertainty as small as 100 m \cite{Carson:2019rjx}. The Square Kilometer Array \citep{SKA} telescope will also play an important role in constraining the possible scenarios for the EoS of matter within the NS.\\

\begin{figure}[!htb]
    \centering
    \includegraphics[width=1.0\linewidth]{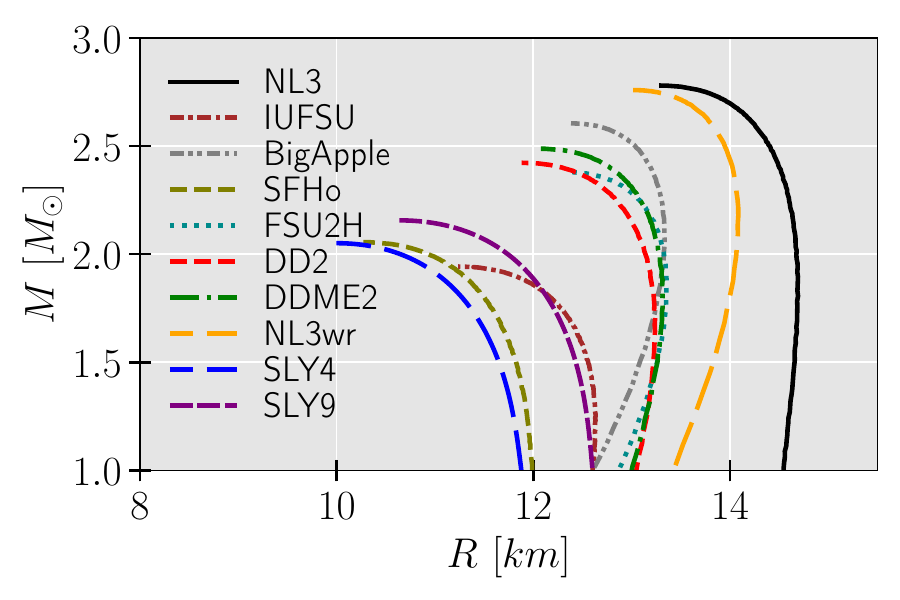}
    \caption{$M(R)$ sequences for nuclear models discussed in the Introduction.}
    \label{fig:0}
\end{figure}

Several microscopic phenomenological models have been used to determine the mass-radius curve that links the astrophysical observations of the NS to the EoS of baryon matter. One of the features that characterise these curves is the slope of the $M(R)$ curve with respect to radius. For masses above 1$M_\odot$, the slope is always negative for some models such as SLy4 and SLy9 \cite{Chabanat:1997un}, SFHo \cite{Steiner:2012xt} and FSU or FSU2 \cite{Chen:2014sca}, see \cite{Douchin:2001sv,Fortin2016,Fortin:2017dsj}, other models such as DD2 \cite{Typel:2009sy}, DDME2 \cite{Lalazissis2005}, NL3$\omega\rho$ \cite{Horowitz:2000xj,Pais:2016xiu}, FSU2H \cite{Tolos:2017lgv} or BigApple \cite{Fattoyev:2020cws} show a backbending and part of the curve above 1$M_\odot$ has a positive slope, see \cite{Fortin2016}, or models like NL3 \cite{Lalazissis:1996rd} or IUFSU \cite{Fattoyev:2010mx} have an almost infinite slope for much of the $M(R)$ curve. The mass-radius curves corresponding to all these models are shown in Fig. \ref{fig:0}.  For the models with a positive or infinite slope, the curve will pass through a negative slope for sufficiently large masses  before the maximum mass star is reached.  

The different behavior of these models can be traced back to the properties of the underlying nuclear interaction. For example, NL3 and NL3$\omega\rho$ have exactly the same symmetric nuclear matter properties, differing only in symmetry energy, with NL3$\omega\rho$ having both a smaller symmetry energy and a smaller slope at saturation. As a consequence, low and medium mass NL3$\omega\rho$ stars have smaller radii than NL3 stars, leading to a backbending of the MR curve. A similar situation occurs when comparing FSU2 and FSU2H, although they do not have the same symmetric nuclear matter, they are similar. The BigApple model, like FSU2H, has a rather small symmetry energy and respective slope at saturation, but its EOS is particularly stiff at high densities, allowing the model to describe a 2.6$M_\odot$ star, thus predicting small radii for low and middle mass stars and large radii for the massive stars. Both FSU and FSU2 have a soft isoscalar behavior at all densities, and this is the main reason for a negative derivative for all masses above 1$M_\odot$. The SFHo EOS was built to produce small radius stars, which is possible with a soft EOS at large densities and a soft symmetry energy. SLy4 and SLy9 have the same behavior as SFHo. EOSs that predict MR curves with an approximately infinite slope for most stars have properties in between the other two sets: NL3 has a hard symmetry energy and a hard nuclear matter EOS, IUFSU was obtained from FSU, making the EOS stiffer at high densities.
The behavior of the $M(R)$ curve reflects the properties of the hadronic interaction, with the back-bending being generally associated with a stiffening of the EoS. 

For the models with a positive or infinite slope, the curve will  always pass through a negative slope for sufficiently large masses before the maximum mass star is reached. This  transition to a negative slope is due to a competition between the strong force and the gravitational force.  The onset of new degrees of freedom such as hyperons or quark deconfinement may favor this transition at smaller masses, see Fig. 4 of \cite{Fortin:2016hny} and Fig. 4 of \cite{Ferreira:2020kvu}. Stiffening of the EOS at high densities has also been predicted by the onset of a quarkyonic phase, \cite{McLerran:2018hbz,Zhao:2020dvu}. The onset of this phase, contrary to the effect of the onset of a quark phase or hyperonic degrees of freedom, creates a backbending in the $M(R)$ curve, see Fig. 4 of reference \cite{McLerran:2018hbz}.

Recently, the authors of \cite{Ecker:2024eyf} selected six EoS that maximize the variance of four NS properties from their dataset of model-independent EoS constructed using Gaussian processes. Of the six EoS, five predict $M(R)$ curves with a positive derivative between 1.0$M_\odot$ and $M\gtrsim1.7M_\odot$, and the sixth has a negative slope for all stars with masses above 1.0$M_\odot$, a behaviour similar to that of SLy4 and SFHo. Since the behaviour of the $M(R)$ curve is inextricably linked to the density dependence of the EoS, it makes sense to ask whether the constraints imposed to build the data sets require the $M(R)$ curves to have a positive slope in some range of NS masses, in which mass range, and what is the reason for this behaviour. Knowledge of this behaviour from observations would place strong constraints on the acceptable microscopic models.

In the present work, using a model-independent, agnostic description of NS matter, we aim to answer the question: is it possible to identify NS properties that distinguish the two sets of models that do or do not have a negative slope in the whole mass range above 1$M_\odot$? Note that if sufficient precision is achieved in the determination of the NS radius, back-bending will be easily confirmed if a low-mass star has a smaller radius than a medium to high-mass star. Knowing the behaviour of the slope of $M(R)$ would place strong constraints on the microscopic model and would provide information on the high density behaviour of the QCD EoS.\\ 

The paper is organized as follows. The parametrization we use to describe neutron star matter is presented in Sec.~\ref{sec:dataset}. The results are shown and discussed in Sec.~\ref{sec:results} and  some  conclusions are drawn in Sec.~\ref{sec:conclusions}.

\section{\label{sec:dataset} Dataset}

To describe the equation of state (EoS) of neutron star matter, we use the piecewise polytropic parameterization $p(\rho)=K\rho^{\Gamma}$, where $\rho=m\, n$ is the rest mass density, $n$ is the baryon number density, $m$ is the mass of a baryon,  $K$ is the polytropic pressure coefficient, and $\Gamma$ is the adiabatic index \cite{Read:2008iy,Hebeler:2013nza}. 
The present parameterization uses five connected polytropic segments. The first segment is defined within the density range $[n_{\text{crust}},1.1n_0]$, where $n_0=0.16$ fm$^{-3}$ denotes the nuclear saturation density and $n_{\text{crust}}\equiv n_0/2$, with the polytropic index $\Gamma_0$ randomly chosen from the range $1.0<\Gamma_0<4.5$. This $\Gamma_0$ interval ensures that the polytrope lies within the band described by the chiral effective field theory \cite{Hebeler:2013nza}. We assume the SLy4 EoS \cite{sly4} for densities $n<n_{\rm crust}$ .
To obtain a flexible and robust representation of neutron star EoS, the remaining four polytropic segments start at random densities, such that $n_1<n_2<n_3<n_4$, with randomly chosen polytropic indices $\chav{\Gamma_1,\Gamma_2,\Gamma_3,\Gamma_4}$. The parameter space $\chav{\Gamma_0,\Gamma_1,n_1,\Gamma_2,n_2,\Gamma_3,n_3,\Gamma_4,n_4 }$ was covered by randomly sampling from uniform distributions consistent with $\chav{n_1,n_2,n_3,n_4}$ being within $n_0$ and $8n_0$, $1. 0<\Gamma_0<4.5$ and $0.05<\Gamma_i<8$ for $i=1,...,4$. 

It has been discussed that uncertainties are introduced in the calculation of the radius of low or medium mass stars if a unified inner crust is not considered, while the outer crust has a negligible influence on the determination of the radius \cite{Fortin2016,Pais:2016xiu}.    In \cite{Pais:2016xiu} it was shown that the uncertainty introduced in the determination of the radius $R$ is small when the inner crust used has been calculated within a model with a similar symmetry energy slope, in particular for masses of the order of 1.4$M_\odot$, $\Delta R \lesssim$ 1\% \cite{Pais:2016xiu}. Imposing the constraints from $\chi$EFT strongly limits the symmetry energy, and the slope $L$ takes values in the range $35\lesssim L\lesssim 65$ MeV \cite{Huth:2020ozf}.  The symmetry energy of SLy4 at saturation is 46 MeV, so it is expected that the uncertainty introduced in the present calculation of the radius is not large.  In \cite{Malik:2024nva}, with a slightly different treatment of the crust but also imposing the $\chi$EFT conditions, the radius of a 1.4$M_\odot$ star obtained with a unified EOS did not differ by more than 1.5\% (corresponding to $\sim 200$ m), and in most cases only half this amount. Taking this uncertainty into account, the limits given in the following sections should be considered as indicative.  Note that it has recently been shown that it is also possible to constrain the crust-core transition and the symmetry energy using asteroseismic constraints \cite{Neill:2022psd,Sorensen:2023zkk}.

To study NS properties, we solve the Tolmann-Oppenheimer-Volkoff (TOV) equations \cite{1939PhRv...55..374O,1939PhRv...55..364T}, which describe spherically symmetric stars in hydrostatic equilibrium. Furthermore, the differential equations that determine the tidal deformability of the stars have also been solved \cite{Hinderer:2009ca}. 
A valid EoS must be consistent with the observation of a $M>2M_{\odot}$ NS and have a speed of sound that remains less than the speed of light. We have generated a data set containing 40435 valid EoS.

Depending on the value of the slope $dM/dR$ calculated along the TOV $M(R)$  sequence between $1.0M_{\odot}$ and $M_{\mathrm{max}}$, we divide the dataset into two subsets: a) $3076$ EoSs whose $M(R)$ sequences satisfy $dM/dR<0$ in the whole mass range above 1$M_\odot$ and b) $37359$ EoSs that do not fulfil $dM/dR<0$.

We considered the following astrophysical constraints (95\% CIs): i)  $10.71\text{ km}<R(2.07M_{\odot})<15.02\text{ km}$ \cite{2021ApJ...918L..27R} and $11.14\text{ km}<R(2.06M_{\odot})<20.20\text{ km}$ \cite{2021ApJ...918L..28M} for PSR J0740+6620; and ii)
$10.94\text{ km}<R(1.44M_{\odot})<15.50\text{ km}$ \cite{2019ApJ...887L..24M} and $10.57\text{ km}<R(1.34M_{\odot})<14.86\text{ km}$ \cite{2019ApJ...887L..21R} for PSR J0030+0451. 
The PSR J0740+6620 observations were implemented by imposing $R(2.0M_{\odot})>10.71\text{ km}$. This is less restricitive than $R(2.07M_{\odot})>10.71\text{ km}$ but takes into account also the uncertainty associated with the determination of the PSR J0740+6620 mass.
Furthermore, the effective tidal deformability of $\tilde{\Lambda}<720$ (low spin-prior), where $\tilde{\Lambda}=(16/13)((12q+1)\Lambda_1+(12+q)q^4\Lambda_2)/(1+q)^5)$, estimated from the GW170817 event \cite{Abbott:2018wiz}, with a binary mass ratio of $0.73\leq q=M_2/M_1 \leq1$ and a chirp mass of  $1.186M_{\odot}$, where $M_{\text{chirp}}=(M_1M_2)^{3/5}/(M_1+M_2)^{1/5}
$. For a review of theoretical, experimental and observational constraints for the equation of state of dense  matter see the recent review \cite{MUSES:2023hyz}. After applying the above  constraints we ended up with a total of 21736 EoS: 2493 with $dM/dR<0$ and 19243 with $dM/dR\nless 0$. As the astrophysical constraints were applied via hard cutoffs, the present study should be seen as an approximation while a more rigorous procedure would require a full likelihood analysis.

\section{\label{sec:results} Results}
 
Several studies have been carried out to identify the mass-radius region allowed by current astrophysical observations considering model-independent EOS, including different parameterizations of the EOS such as piecewise polytropic, speed-of-sound or spectral interpolations, or  non-parametric approaches  \cite{Annala:2017llu,Landry:2020vaw,Annala2019,Raaijmakers2021,Annala:2021gom,Altiparmak:2022bke}. Similar studies were performed considering  nuclear metamodels, such as \cite{Thi:2021jhz}, or 
microscopic phenomenological nuclear  models, see \cite{Traversi_2020,Malik:2022zol,Providencia2024} among others. In the present study we consider the piecewise polytropic interpolation \cite{Annala:2017llu}. 

\begin{figure}[!htb]
    \centering
    \includegraphics[width=1.0\linewidth]{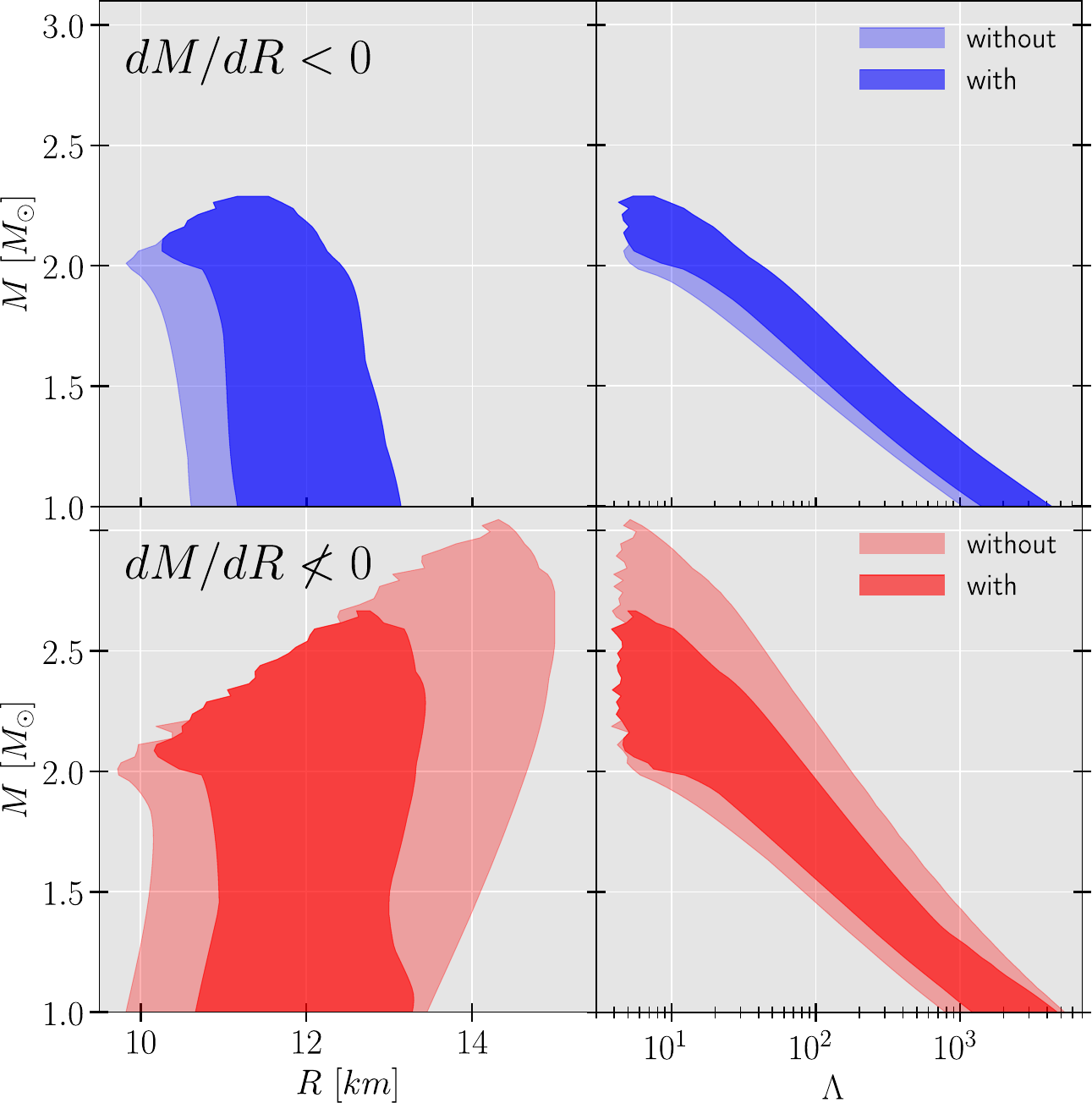}
    \caption{Sequences $M(R)$ (left) and $M(\Lambda)$ (right) for the set of EoS satisfying $dM/dR<0$ (top) and $dM/dR \nless 0$ (bottom) with (dark) and without (light) astrophysical constraints (see text for details). Maximum masses vary between 2.01$\lesssim M_{\text{max}}/M_\odot \lesssim 2.20$ at 90\% CI for $dM/dR<0$ even if  astrophysical constraints are imposed, and 2.01$\lesssim M_{\text{max}}/M_\odot \lesssim 2.43$ (2.01$\lesssim M_{\text{max}}/M_\odot \lesssim 2.60$)  for $dM/dR\nless0$, imposing (not imposing) astrophysical constraints.  The boundaries of the different regions specify the extremes (minimum/maximum).}
    \label{fig:MR}
\end{figure}

The $M(R)$ and $M(\Lambda)$ sequences are shown in Fig.~\ref{fig:MR} for both subsets with (dark colors) and without (light colors) astrophysical constraints. Some comments are in order: i) EoS which satisfy  satisfy $dM/dR<0$, predict smaller maximum masses, and smaller upper bounds for the NS radii. Medium mass stars with a radius above 13 km or a maximum mass above $2.3M_\odot$ would exclude this set of EoS. Astrophysical constraints filter $M(R)$ curves with a small radius; ii) the set of EoS that show a portion with positive slope, allows for stars as massive as 3$M_\odot$, and a and radii above 14 km, if the astrophysical conditions are not imposed. These values are reduced, respectively, to  maximum masses $\sim$2.6$M_\odot$  and radii below $\sim 13.5$ km. Astrophysical constraints also cut the low radii $M(R)$ curves.

\subsection{\label{sec:results_part1} Neutron star equation of state}

\begin{figure}[!htb]
    \centering
    \includegraphics[width=1.0\linewidth]{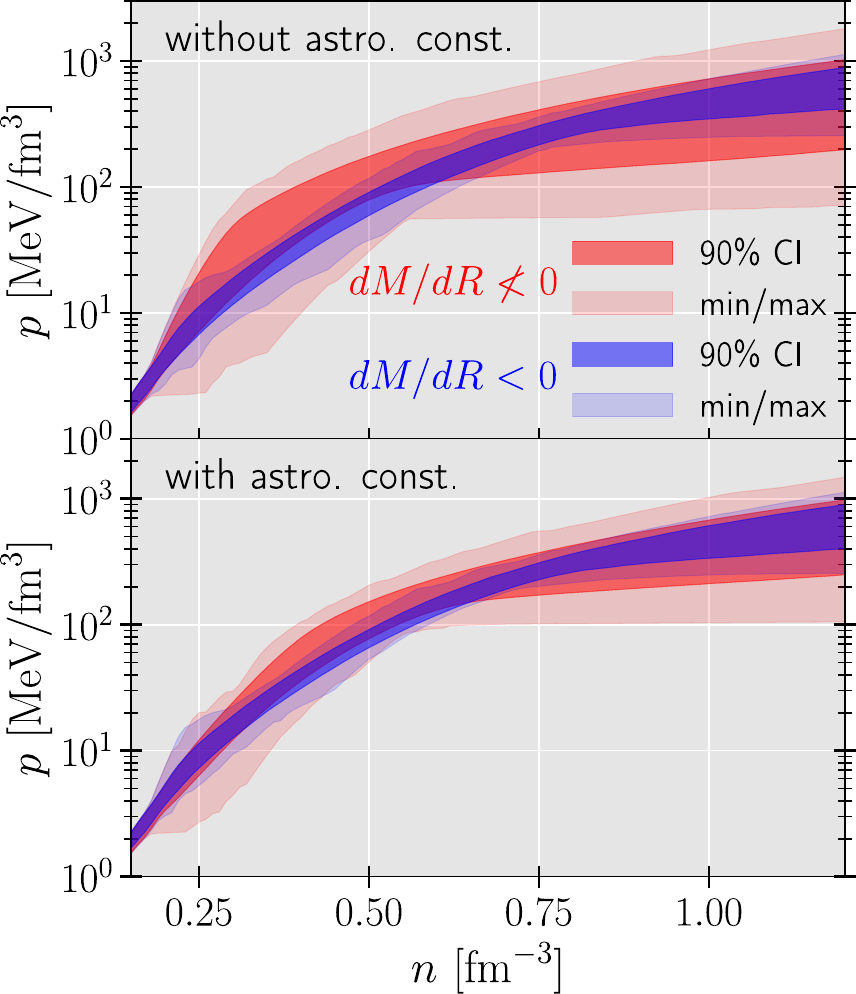}
    \caption{Pressure as a function of baryon density for the set $dM/dR<0$ (blue) and $dM/dR\nless 0$ (red) with (bottom) and without (top) astrophysical constraints. The darker bands represent 90\% credible intervals (CI) while the lighter bands show the extremes (minimum/maximum).  }
    \label{fig:1}
\end{figure}

The constraint that a negative slope along the entire $M(R)$ sequence (no backbending) has on the pressure of neutron star matter is shown in Fig.~\ref{fig:1}, where we compare the $p(n)$ curves for both sets. The dark colours represent the 90\% credible intervals (CI), while the light colours define the full data set, i.e. the extremes.  The pressure has a much smoother behaviour for the $dM/dR<0$ set, being constrained to much lower values (softening) at $n\approx0.4-0.6$ fm$^{-3}$: the 90\% CI of the $dM/dR<0$ is below the 90\% CI of the red set. {However, note that at high density this set is on average stiffer than the set with backbending. This is due to the condition that the maximum mass must be at least 2$M_\odot$}. The kink seen in the statistics of the red set at about $n=0.55$ fm$^{-3}$, which has been interpreted as a possible indication of a phase transition, is not present in the blue set.  The imposition of the astrophysical constraints reduces the range covered, but the main features described above are still present. 

\begin{figure}[!htb]
    \centering
    \includegraphics[width=1.0\linewidth]{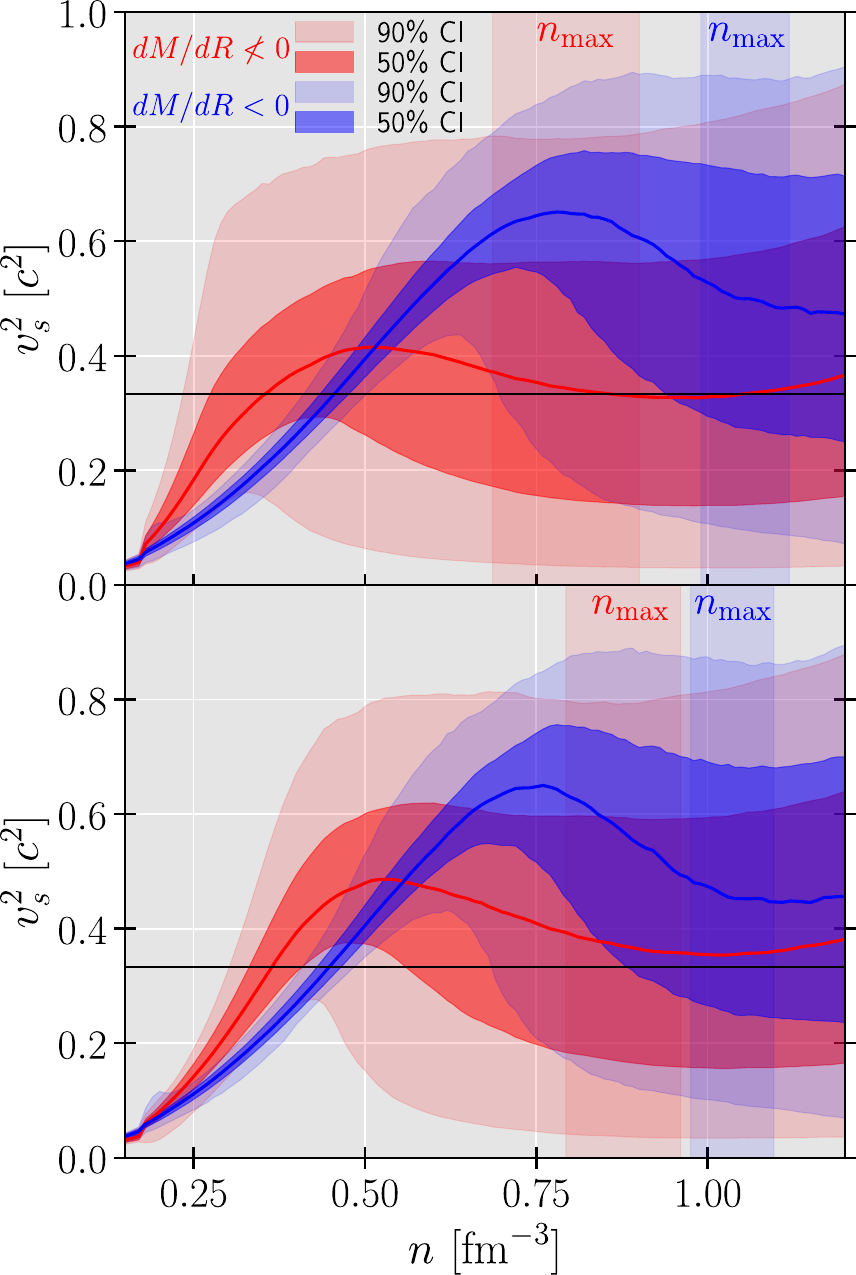}
    \caption{Speed-of-sound
squared as a function of baryon density for the set $dM/dR<0$ (blue) and $dM/dR\nless 0$ (red) with (bottom) and without (top) astrophysical constraints. The lighter bands represent 90\% CI, the darker bands 50\% CI, and the solid lines the median. The vertical bands specify the 65\% CI for the central densities ($n_{\mathrm{max}}$) at the $M_{\mathrm{max}}$. 
}
    \label{fig:2}
\end{figure}

Figure \ref{fig:2} shows the speed-of-sound
squared for both sets.  We also show the 65\% vertical bands representing the central densities of $M_{\mathrm{max}}$ for both sets. The blue set ($dM/dR<0$, no backbending on the $M(R)$ diagram) has a $v_s^2(n)$ that increases almost monotonically up to $\sim 0.8$ fm$^{-3}$, the median (solid lines) has a bend at this density, before decreasing up to the density $n_{\mathrm{max}}$, but does not reach the conformal limit $v_s^2/c^2=1/3$. On the other hand, the red set has a median value close to $v_s^2/c^2=1/3$ at the central densities. The astrophysical constraints originate the peak around 3$n_0$ as identified in other works \cite{Annala2019,Altiparmak:2022bke}. These astrophysical conditions remove from the dataset the hardest EoS at the lowest densities. There is a clear distinction that can be made from the assessment of possible backbending in the $M(R)$ plot.  The $dM/dR<0$ set has larger central densities because, as the EoS is softer, $v_s^2$($dM/dR<0$) remains below $v_s^2$($dM/dR\nless0$) for $n\lesssim 3 n_0$, matter is compressed more efficiently by gravity. Note also that the position of the central densities in the case of $dM/dR\not<0$ ($dM/dR<0$) shifts to larger (smaller) densities because some of the hardest (softest) EoS have been removed from the set.

\begin{figure}[!htb]
    \centering
    \includegraphics[width=1.0\linewidth]{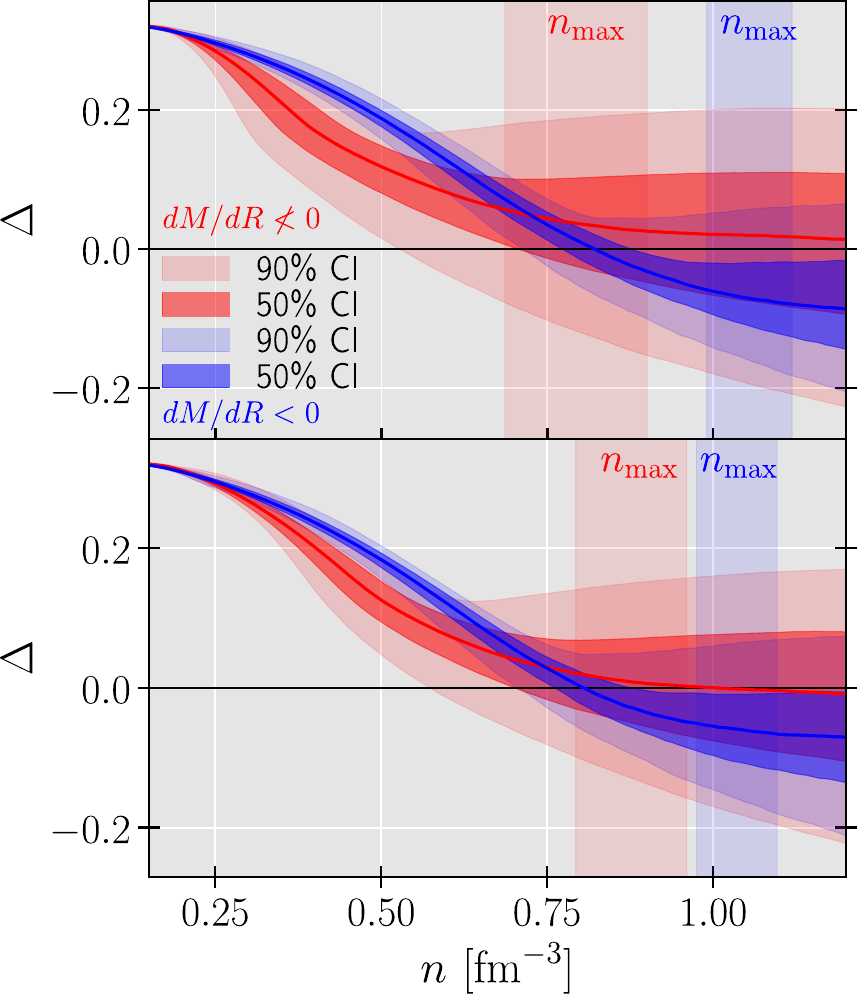}
    \caption{The normalized matter trace anomaly as a function of baryon density for the set $dM/dR<0$ (blue) and $dM/dR\nless 0$ (red) with (bottom) and without (top) astrophysical constraints. The lighter bands represent 90\% CI, the darker bands 50\% CI, and the solid lines the median. The vertical bands specify the 65\% CI for the central densities ($n_{\mathrm{max}}$) at the $M_{\mathrm{max}}$.}
    \label{fig:3}
\end{figure}

In \cite{Fujimoto:2022ohj} the authors consider the renormalized trace anomaly $\Delta=1/3-p/e$, where $p$ is the pressure and $e$ is the energy density, to study conformality restoration within NS. Since perturbative QCD predicts a small positive trace in the very high density limit, it was conjectured that the renormalized trace anomaly should be positive in the whole density range from low to high density.  The anomaly of the normalized matter trace is shown in Fig. \ref{fig:3} for both sets. For the $dM/dR<0$ set, the 50\% CI crosses the zero axis and remains below zero. However, the other set of EoS   has a positive $\Delta$ above the median throughout the density range. The conjecture proposed in \cite{Fujimoto:2022ohj} seams to be lightly favoured by a $M(R)$ curve with backbending by comparing the median values.

\begin{figure}[!htb]
    \centering
    \includegraphics[width=1.0\linewidth]{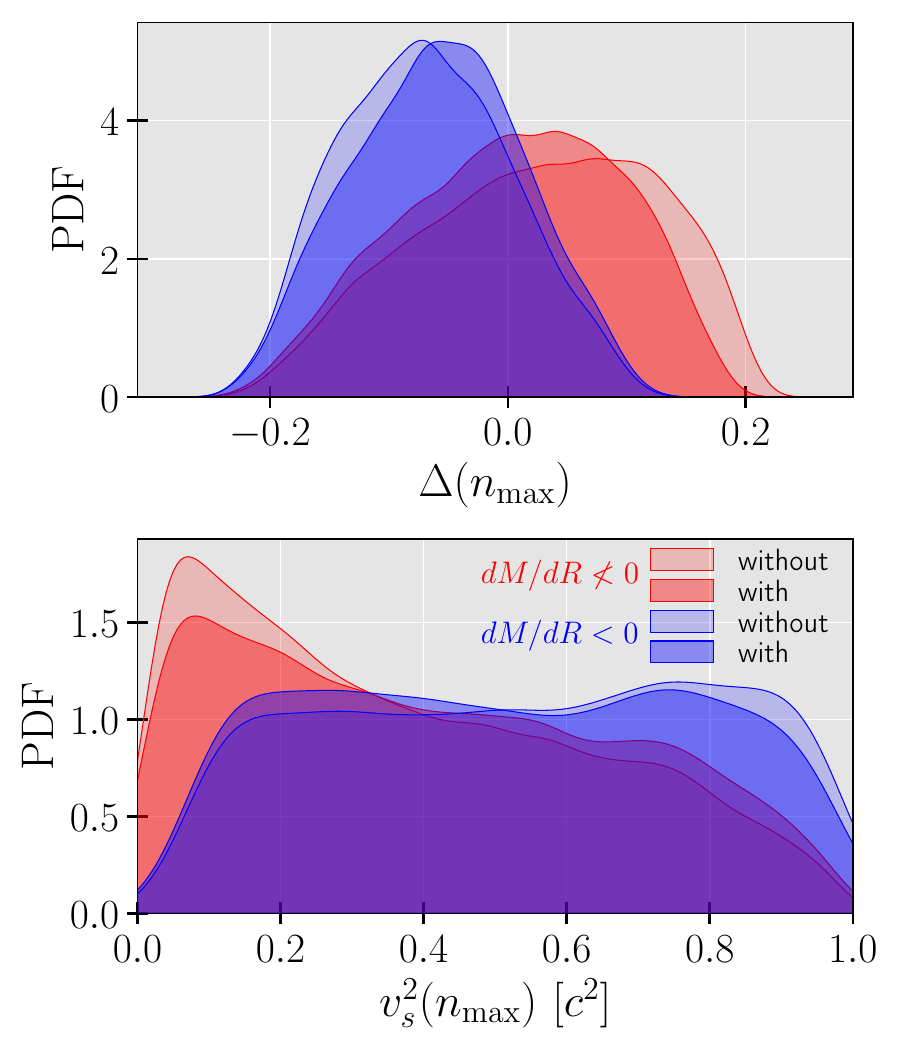}
    \caption{The  probability distribution functions  for the normalized matter trace anomaly (top) and speed-of-sound squared (bottom) at the central densities of $M_{\mathrm{max}}$. We display the sets $dM/dR<0$ (blue) and $dM/dR\nless 0$ (red) with (dark) and without (light) astrophysical constraints. The 50\% CIs for $\Delta(n_{\mathrm{max}})$ are $-0.070^{-0.051}_{+0.053}$ ($dM/dR<0$, without restrictions), $-0.057^{-0.052}_{+0.049}$ ($dM/dR<0$, with restrictions), $0.032^{-0.081}_{+0.073}$ ($dM/dR\nless0$, without restrictions), and $0.008^{-0.073}_{+0.066}$ ($dM/dR\nless0$, with restrictions).
    }
    \label{fig:4}
\end{figure}

A clearer analysis is obtained from the probability distribution functions for both $\Delta$ and $v_s^2$ at the central densities of $M_{\mathrm{max}}$ shown in Fig.~\ref{fig:4}. The speed-of-sound squared probability distributions of the two sets at the centre of the maximum mass star are quite different: the $dM/dR<0$ set has a rather flat profile which is cut off by causality at $v_s^2/c^2=1$; the $dM/dR\nless 0$ set, on the other hand, presents a distribution with a well-defined peak around $v_s^2/c^2=0.1$, which decreases monotonically to $v_s^2/c^2=1$, and the causality condition has a very small effect. The renormalized trace anomaly probability distribution functions calculated at the centre of the maximum mass star also show different behaviour: for the $dM/dR\nless 0$ set, most of the EoS have a positive value, while the $dM/dR< 0$ set shows negative values for a very reduced number of EoS. As expected from Fig~\ref{fig:3}, the 50\% CI for the $dM/dR< 0$ set is always positive, see the caption of Fig.~\ref{fig:4}. 

\begin{figure}[!htb]
    \centering
    \includegraphics[width=1.0\linewidth]{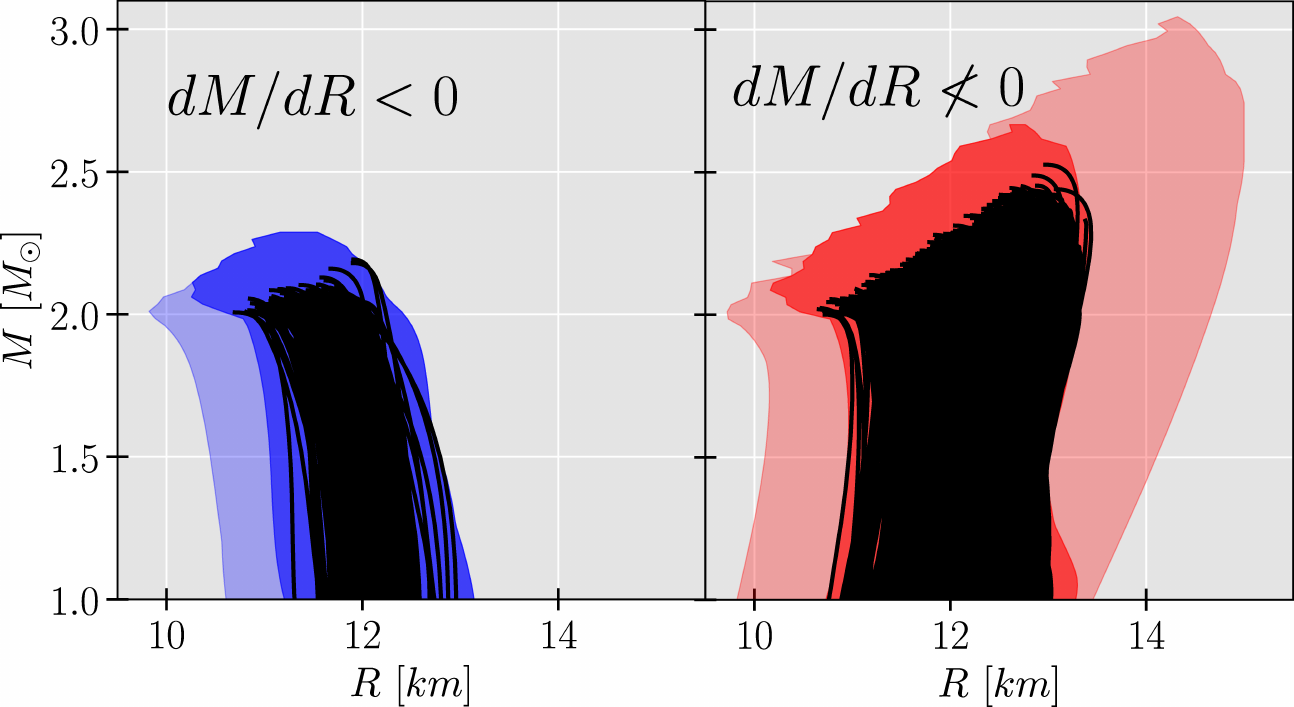}
    \caption{
    Same plots as in Fig.~\ref{fig:MR} but showing explicitly the $M(R)$ sequences for the EOS with positive $\Delta(n)$ for $0<n<1.2$ fm$^{-3}$.  }
    \label{fig:temp}
\end{figure}

To complete this study, we have identified all the EoSs from both sets that have a positive trace anomaly up to $1.2$ fm$^{-3}$ and we have plotted them on top of the $M(R)$ plot, see Fig.~\ref{fig:temp}. 
As discussed in \cite{Fujimoto:2022ohj}, the maximum masses reached by these EoS are lower and have a larger radius for a given mass, regardless of the behaviour of the slope. For the same maximum mass, the difference between the radius of the star with positive and the one with  negative trace anomaly  can be  as large as  1 km for the non-negative slope set  or even larger for the negative slope set.

Tables \ref{tab:1} and \ref{tab:2} show several properties of NS with different masses. In Table \ref{tab:1} the median and 90\% CI limits are given for the analysis with and without additional astrophysical constraints.  In Table \ref{tab:2} some extreme properties are also given.

\subsection{\label{sec:results_part2} Constraining $R$ and $\Lambda$ from $dM/dR$}

Assuming that future observations will allow the estimation of $dM/dR$ around specific neutron star masses, $M_i$, we analyse below what constraints can be extracted on the radius and tidal deformability of NS from the value of $dM/dR|_{M=M_i}$. These derivatives were obtained by interpolating the function $(dM/dR,M)$ at specific NS mass $M_i$, where finite differences was used to estimate $dM/dR$.
Note that the onset of an exotic degree of freedom, such as hyperons or deconfined quark matter, can cause the transition from $dM/dR> 0$ to $dM/dR<0$. It is often predicted that the opening of new degrees of freedom occurs above twice the saturation density, inside stars of mass $\gtrsim 1.4\, M_\odot$. We will therefore compare some properties of the EoS that can be affected by the slope at 1.4$M_\odot$, in particular, the radius of 2$M_\odot$ star and  the radius and mass of maximum mass stars.

\begin{figure}[!htb]
    \centering
    \includegraphics[width=1.0\linewidth]{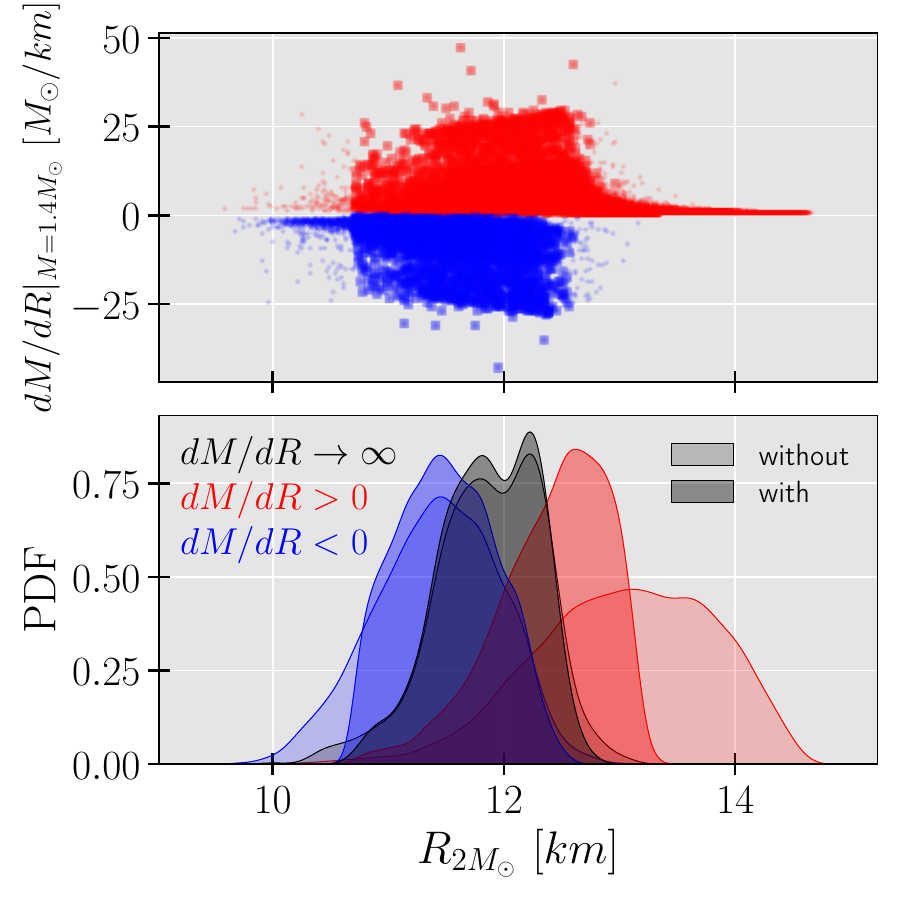}
    \caption{ Scatter plot of $dM/dR|_{M=1.4M_{\odot}}$ vs. $R(2.0M_{\odot})$ (top) and the respective $R(2.0M_{\odot})$ PDF (bottom). The colors indicate the $dM/dR|_{M=1.4M_{\odot}}$ value: negative (blue), positive but finite (red), and  $\infty$ (black) with (dark colors) and without (light colors) astrophysical constraints.}
    \label{fig:5}
\end{figure}

\begin{figure*}[!htb]
    \centering
    \includegraphics[width=0.48\linewidth]{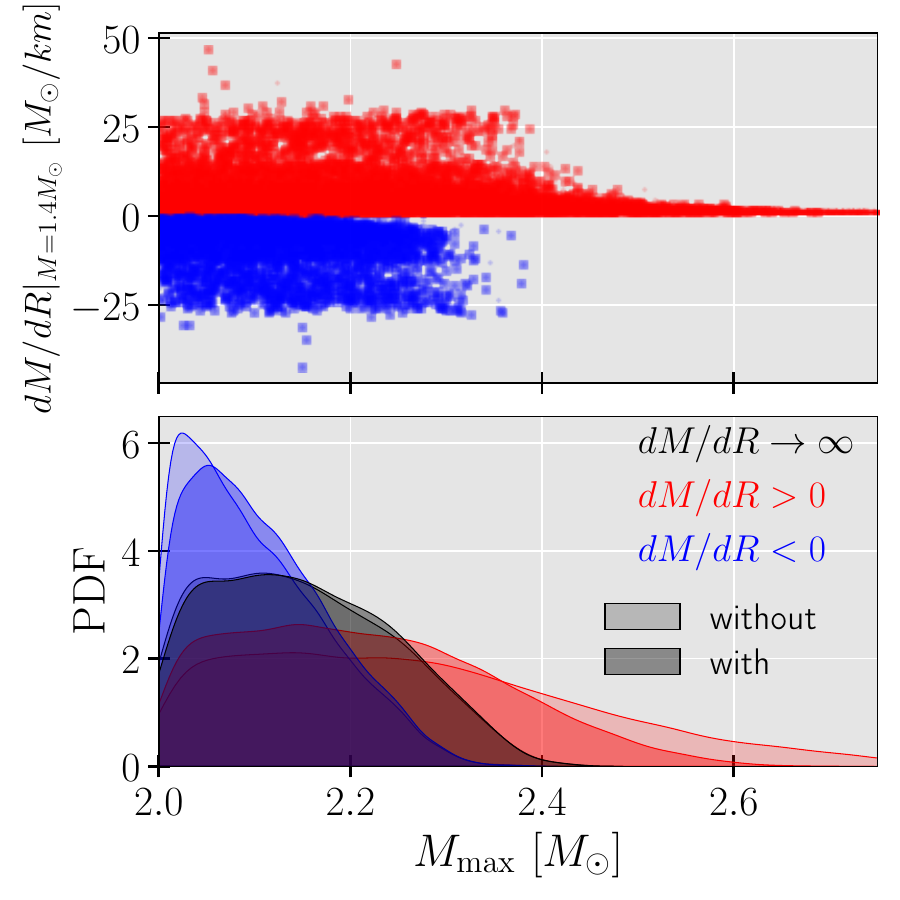}
    \includegraphics[width=0.48\linewidth]{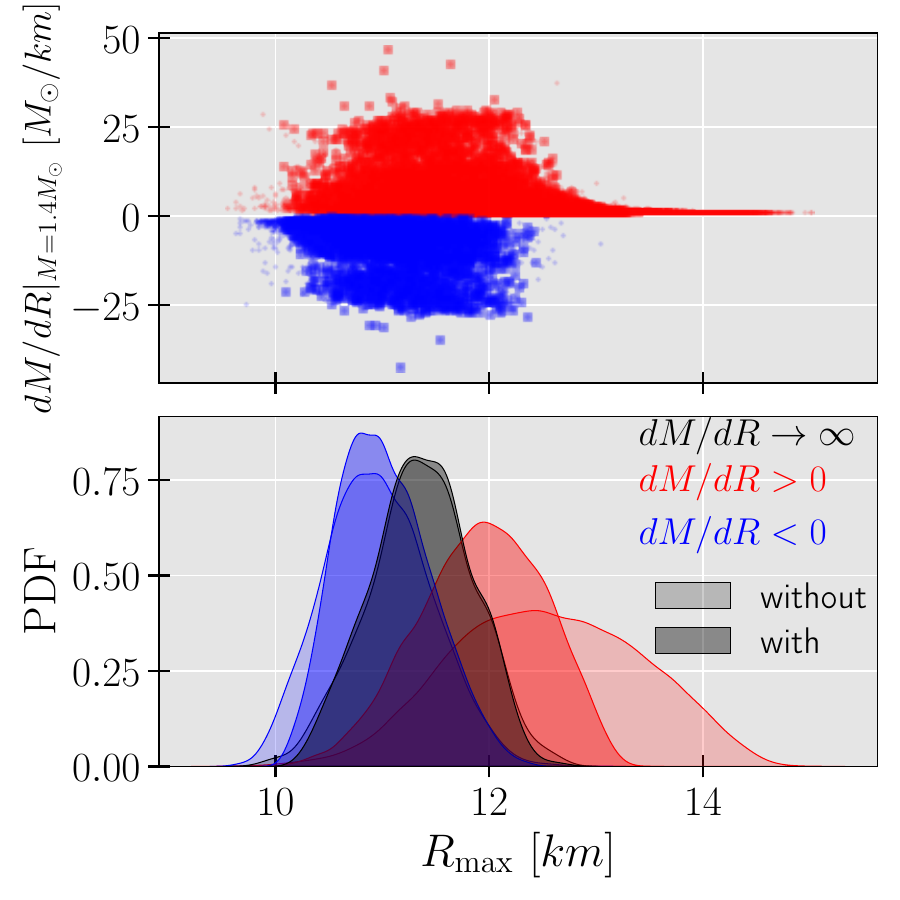}
    \caption{Scatter plot of $dM/dR|_{M=1.4M_{\odot}}$ vs. $\chav{M_{\mathrm{max}}, R_{\mathrm{max}}}$ (top panels) and the respective $\chav{M_{\mathrm{max}}, R_{\mathrm{max}}}$ PDFs (bottom panels). The colors indicate the $dM/dR|_{M=1.4M_{\odot}}$ value: negative (blue), positive but finite (red), and  $\infty$ (black) with (dark colors) and without (light colors) astrophysical constraints. 
    }
    \label{fig:6}
\end{figure*}

\begin{figure}[!htb]
    \centering
    \includegraphics[width=1.0\linewidth]{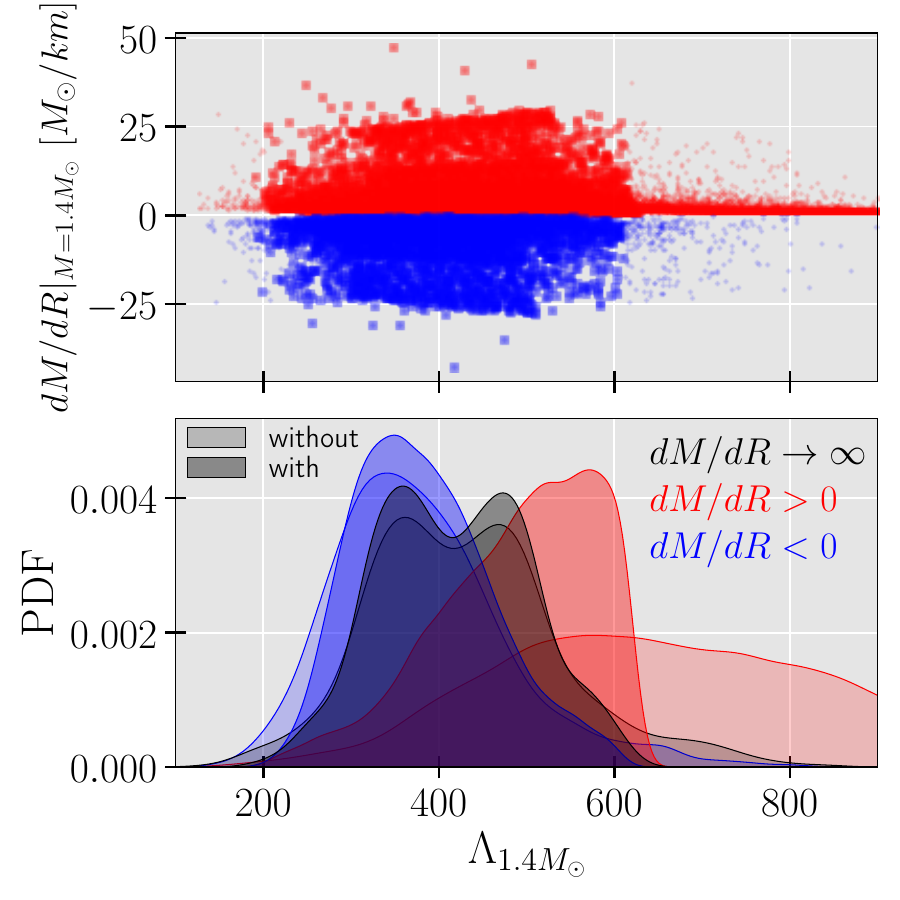}
    \caption{ Scatter plot of $dM/dR|_{M=1.4M_{\odot}}$ vs. $\Lambda(1.4M_{\odot})$ (top) and the respective $\Lambda(1.4M_{\odot})$ PDF (bottom). The colors indicate the $dM/dR|_{M=1.4M_{\odot}}$ value: negative (blue), positive but finite (red), and  $\infty$ (black) with (dark colors) and without (light colors) astrophysical constraints. 
    }
    \label{fig:7}
\end{figure}

\begin{figure*}[!htb]
    \centering
    \includegraphics[width=0.45\linewidth]{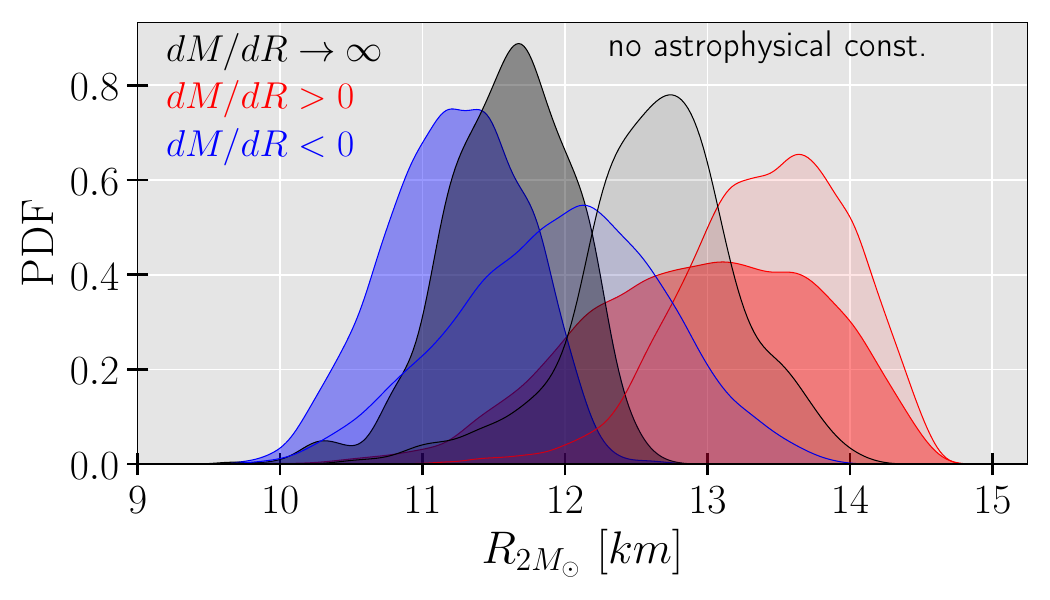}
    \includegraphics[width=0.45\linewidth]{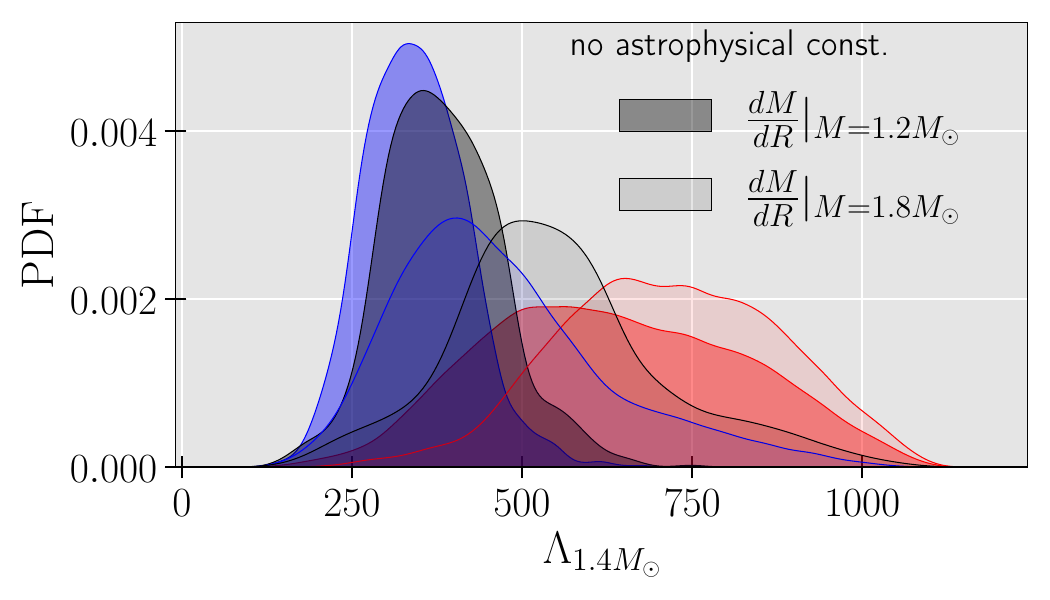}\\
    \includegraphics[width=0.45\linewidth]{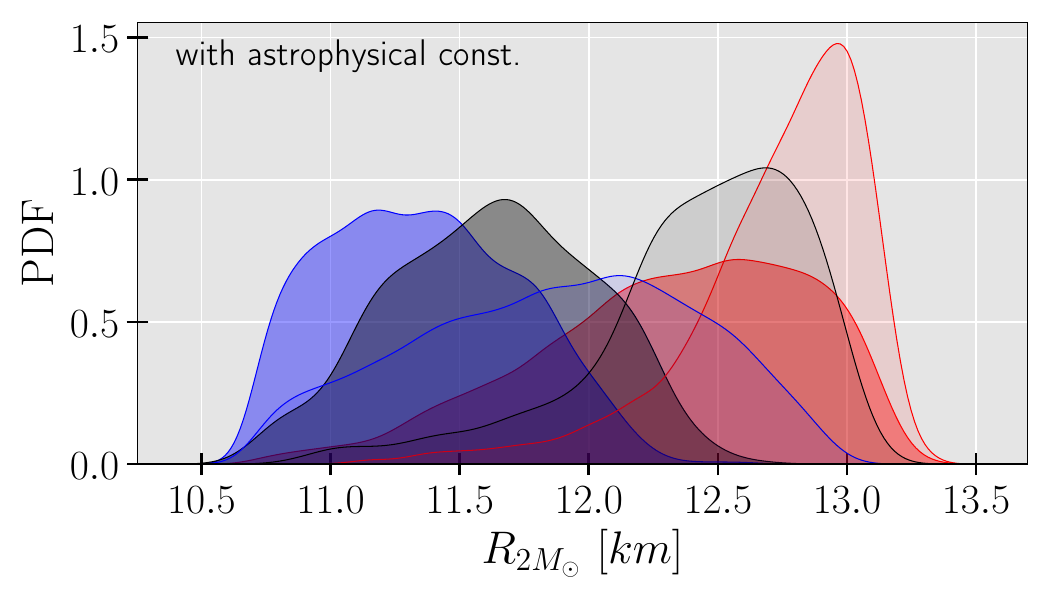}
    \includegraphics[width=0.45\linewidth]{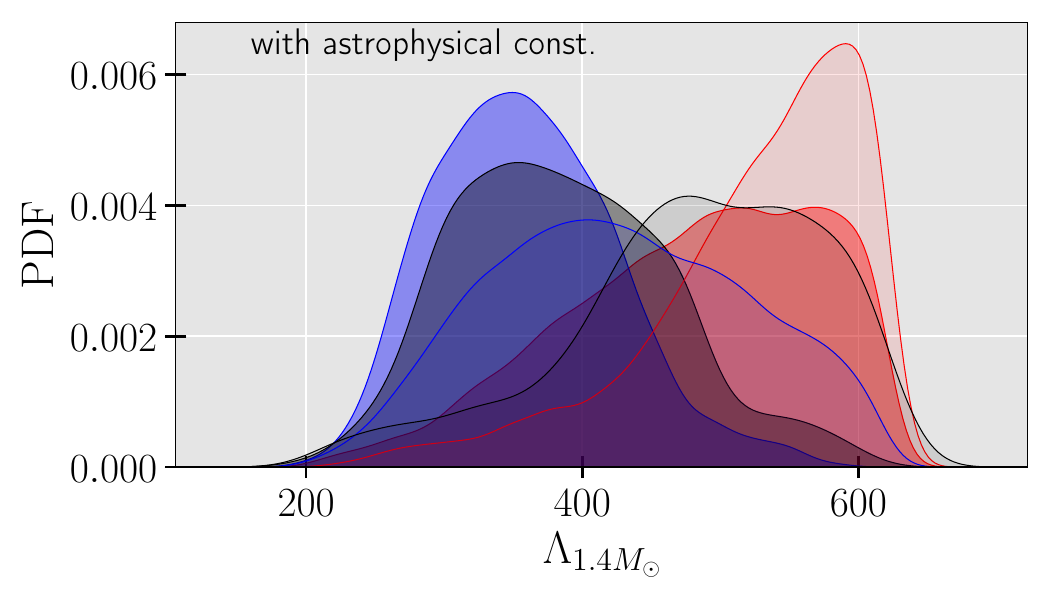}
    \caption{The PDFs of the radius of $2.0M_{\odot}$ (left) and the tidal deformability of a $1.4M_{\odot}$ (right) with (bottom) and without (top) astrophysical constraints applied for $dM/dR|_{M=1.2M_{\odot}}$ (dark colors) and $dM/dR|_{M=1.8M_{\odot}}$ (light colors) values: negative (blue), positive but finite (red), and  $\infty$ (black). }
    \label{fig:8}
\end{figure*}
Figure~\ref{fig:5} displays the relation between the value of $dM/dR$ at $M=1.4M_{\odot}$ and $R_{2.0M_{\odot}}$ (top), and the PDF for $R_{2.0M_{\odot}}$ (bottom) with (dark colors) and without (light colors) astrophysical constraints considering the three possibilities for $dM/dR|_{M=1.4M_{\odot}}$: negative (blue), positive but finite (red), and  $\infty$ (black). The black set represents the EoSs for which the $R$ remains almost constant with increasing $M$. Figures \ref{fig:6} and \ref{fig:7} show the same information but for the maximum mass, $M_{\mathrm{max}}$, and the tidal deformability of a $1.4M_{\odot}$ NS, respectively. {Considering no astrophysical constraints (light colors in Figs. \ref{fig:5} to \ref{fig:7}), we have 32035 EOS in the red set, 7328 in blue set, and 1072 in black set, while considering astrophysical constraints (dark colors in the same figures) the numbers decrease to 14395 EOS in red set, 6239 in blue set, and 980 in black set.} Note that the  number of EoS with $dM/dR<0$ at $1.4M_{\odot}$ is larger than the number of EoS with $dM/dR<0$ over the whole range of masses as expected (if no constraints are imposed we have 7328 in the first set and  3106 in the second). A change of slope close to $1.4M_{\odot}$ may just indicate some softening of the nuclear force at large densities (in a RMF description this could be played by the $\omega^4$ term) or the onset of new degrees of freedom such as hyperons or quarks.

From Fig.~\ref{fig:5}, we conclude that $dM/dR<0$  at 1.4$M_\odot$ implies a 2$M_\odot$ radius below 13 km. A radius above 13 km requires a positive slope at 1.4$M_\odot$, or at least an infinite slope. However, a radius below 13 km does not distinguish the values of $dM/dR$ at $M=1.4M_{\odot}$. From the PDF (bottom panel), we see that a radius of 11 km or below will indicate $dM/dR<0$ or an infinite slope , with a very high probability. An analyzes of Fig.~\ref{fig:6}, which relates the slope of the $M(R)$ curve at 1.4$M_\odot$ with the maximum mass, shows that a positive slope at $1.4M_\odot$ is necessary for the condition $M_{\mathrm{max}}\gtrsim 2.4 M_\odot$ to be satisfied.  Concerning the tidal deformability $\Lambda_{1.4}$, a value below 800 does not exclude any set, but if in the future it is possible to restrict this property to values below 500 there is a high probability that the $M(R)$ curve at 1.4$M_\odot$ does not show a back-bending. Note that astrophysical constraints limit the star maximum mass to values below $\sim$2.6$M_\odot$ and the radius of $2.0M_\odot$ and $M_{\mathrm{max}}$ stars to values below 13.5 km, or $\lesssim 12$ km if $dM/dR<0$ at $M=1.4M_{\odot}$ (for the maximum mass below 11.5km at 90\% CI with 12.12 km the extreme and for $2.0M_{\odot}$ below 11.9 km at 90\% CI). This is very interesting information to constrain the microscopic modeling of the EoS.

Figure \ref{fig:8} displays how local information regarding the sign of the slope of the $M(R)$ sequence  at 1.2$M_\odot$ and 1.8$M_\odot$ translates into constraints on $R_{2.0M_{\odot}}$ and $\Lambda_{1.4M_{\odot}}$.
The most constraining conditions, both for the $R_{2.0M_{\odot}}$  and the  $\Lambda_{1.4M_{\odot}}$, come from an early onset of a negative slope or infinite slope, favoring in the first case radii below 12.5 km and a tidal deformability below 600 and in the second values of the radii below 13 km  and of the tidal deformability below 700. The possible restrictions that could be extracted from observationally estimating $dM/dR$ at both $M/M_\odot=1.2$ and $1.8$ are indicated by the PDFs overlap regions. Considering $dM/dR<0$ at both masses (or $dM/dR\to \infty$ at 1.2$M_\odot$ and $dM/dR<0$ at 1.8$M_\odot$) result in $10.5\lesssim R_{2.0M_{\odot}}/\text{km}\lesssim12.5$  (or $10.5\lesssim R_{2.0M_{\odot}}/\text{km}\lesssim13$).
The more exotic scenario where $dM/dR\to \infty$ at both masses, i.e., the radius is constant in a wide range of masses,  a stronger constraint of  $11 \lesssim R_{2.0M_{\odot}}/\text{km}\lesssim 12.7$  is obtained. We also conclude that  the astrophysical constraints impose $\Lambda_{1.4M_{\odot}}\gtrsim 200$, in accordance with \cite{Radice:2018ozg}.

\section{\label{sec:conclusions}Conclusions}

Considering a model-independent set of EoS constructed from five segments of polytropes, we have analysed the information that can be extracted from the slope of the $M(R)$ curve. The set of EoS was constrained by the neutron matter EoS obtained from a chiral effective field at low densities, the description of a 2$M_\odot$ star, and several observational constraints, in particular the effective tidal deformability obtained from the GW170817 detection and the lower radius limits estimated by NICER for pulsars PSR J0740+6620 and PSR J0030+0451.

We have concluded that an EoS characterized by a negative slope throughout the mass range is quite restrictive: the maximum mass predicted is 2.20$M_\odot$ with a radius greater than 10.3 km at 90\% CI. If this constraint is relaxed, the maximum mass increases to 2.43$M_\odot$ at 90\% CI and can reach $\sim 2.69\, M_\odot$ and a maximum radius $\lesssim13.4$km. Note that this maximum mass has also been obtained in \cite{Ferreira:2021pni} using a different description for the EoS. A negative slope also predicts a different behavior of the speed of sound, showing a peak at 4-5$n_0$ instead of 3$n_0$ as often obtained in different studies \cite{Annala2019,Altiparmak:2022bke}. The conjecture proposed in \cite{Fujimoto:2022ohj} concerning the renormalized trace anomaly, which has a positive value at all densities, favors the EoS with back-bending, and as discussed in that work, the maximum masses obtained are smaller and, for a given maximum mass, the maximum star has a larger radius.

A rough estimate of the slope of $M(R)$ at two points, which allows the sign to be extracted, is already informative about neutron star observables such as radius and tidal deformability.  We have analysed the consequences of the value the slope takes at $1.4\,M_\odot$ and concluded that a negative slope indicates $R(2M_\odot)\lesssim 11.9$ km and $R(M_{\mathrm{max}})\lesssim 11.5$ km at 90\% CI. The simultaneous determination of the slope at two different masses can also provide additional constraints: for an EoS with a negative slope at both masses, a star of 2$M_\odot$ mass is expected to have a radius in the range $10.5\lesssim R_{2.0M_{\odot}}\lesssim12.5$ km. We conclude by pointing out that knowing the slope of the $M(R)$ curve at a given mass, or at more than one mass, will provide important information for constraining the microscopic EoS of hadronic matter at high densities.

\section*{ACKNOWLEDGMENTS} 
This work was partially supported by national funds from FCT (Fundação para a Ciência e a Tecnologia, I.P, Portugal) under the projects 2022.06460.PTDC with the  DOI identifier 10.54499/2022.06460.PTDC, and 
UIDB/04564/2020 and UIDP/04564/2020, with DOI identifiers 10.54499/UIDB/04564/2020 and 10.54499/UIDP/04564/2020, respectively.

\onecolumngrid
\appendix

\section{Datasets statistical summary\label{ap:stats}}

\begin{table*}[h]
\caption{Percentiles (5th, 50th, and 95th) for some NS and EoS properties.  
The following astrophysical constraints  were considered: i)   $R(2.0M_{\odot})>10.71\text{ km}$ for J0740+6620  \cite{2021ApJ...918L..27R,2021ApJ...918L..28M}); ii)
$10.94\text{ km}<R(1.44M_{\odot})<15.50\text{ km}$ \cite{2019ApJ...887L..24M} and $10.57\text{ km}<R(1.34M_{\odot})<14.86\text{ km}$ \cite{2019ApJ...887L..21R} for PSR J0030+0451; and iii) $\tilde{\Lambda}<720$ for the GW170817 event \cite{Abbott:2018wiz}.
}
\label{tab:1}
\setlength{\tabcolsep}{4.5pt}
      \renewcommand{\arraystretch}{1.4}
\centering
\resizebox{\columnwidth}{!}{%
\begin{tabular}{c ccccccc c ccccccc }
\hline \hline 
&  \multicolumn{7}{c}{Set $dM/dR<0$} & &  \multicolumn{7}{c}{Set $dM/dR\nless0$} \\
\cline{2-8} \cline{10-16} 
& \multicolumn{3}{c}{Without constraints}  & &\multicolumn{3}{c}{With constraints} & &\multicolumn{3}{c}{Without constraints}  & &\multicolumn{3}{c}{With constraints}  \\
\cline{2-4} \cline{6-8} \cline{10-12} \cline{14-16} 
& 5\% & 50\% & 95\% & & 5\% & 50\% & 95\% & & 5\% & 50\% & 95\% & & 5\% & 50\% & 95\%\\ 
 \hline
 $M_{\mathrm{max}}$ [$M_{\odot}$]  &2.01&2.06&2.19 & &2.01&2.08&2.20 & &2.02&2.22&2.60 & &2.02&2.17&2.43  \\
$R_{\mathrm{max}}$ [km]  &10.08&10.69&11.44 & &10.30&10.80&11.49 & &10.81&12.33&13.92 & &10.70&11.74&12.78  \\
$n_{\mathrm{max}}$ [fm$^{-3}$] &0.856&1.057&1.187 & &0.839&1.041&1.152 & &0.571&0.789&1.047 & &0.682&0.873&1.073  \\
$v_s^2(n_{\mathrm{max}})$ [$c^2$] &0.117&0.547&0.944 & &0.114&0.509&0.930 & &0.034&0.326&0.847 & &0.039&0.375&0.869  \\
$p(n_{\mathrm{max}})$ [MeV/fm$^{3}$] &313.679&529.452&765.946 & &307.047&502.103&740.218 & &109.545&289.643&585.372 & &165.500&344.398&621.43  \\
$\Delta(n_{\mathrm{max}})$ &-0.178&-0.070&0.054 & &-0.175&-0.057&0.061 & &-0.145&0.032&0.173 & &-0.152&0.008&1.141  \\
$\Lambda(1.36M_{\odot})$ & 270 &397 &519 & & 312 &417 &526 & &385&711 &1089 & &359&566&710\\
$R(1.4M_{\odot})$ [km] &11.23&11.90&12.40 & &11.46&11.99&12.42 & &11.76&12.86&13.66 & &11.63&12.45&12.86  \\
$n_{\mathrm{max}}(1.4M_{\odot})$ [fm$^{-3}$] &0.450&0.509&0.587 & &0.448&0.497&0.549 & &0.276&0.359&0.489 & &0.354&0.410&0.502  \\
$\Lambda(1.4M_{\odot})$  &223&329&432 & &259&346&438 & &323&607&943 & &302&479&606  \\
$v_s^2(1.4M_{\odot})$ [$c^2$]& 0.350&0.408&0.565 & &0.348&0.403&0.551 & &0.281&0.469&0.641& &0.317&0.450&0.626\\
$R(1.6M_{\odot})$ [km] & 11.10&11.77&12.30&&11.37&11.87&12.32&&11.74&12.94&13.86&&11.61&12.48&11.95\\
$n_{\mathrm{max}}(1.6M_{\odot})$ [fm$^{-3}$] & 0.504&0.574&0.662&&0.500&0.562&0.618&&0.289&0.390&0.544&&0.377&0.449&0.559\\
$R(2.0M_{\odot})$ [km] &10.34&11.14&11.86 & &10.77&11.26&11.89 & &11.35&12.90&14.16 & &11.21&12.29&13.04  \\
$n_{\mathrm{max}}(2.0M_{\odot})$ [fm$^{-3}$] &0.659&0.819&1.043 & &0.652&0.819&0.940 & &0.328&0.491&0.789 & &0.438&0.577&0.820  \\
\hline
\hline

\end{tabular}
}
\end{table*}

\begin{table*}[h]
\caption{Extremes (maximum/minimum) for some NS and EoS properties. The following astrophysical constraints  were considered: i)   $R(2.0M_{\odot})>10.71\text{ km}$ for J0740+6620  \cite{2021ApJ...918L..27R,2021ApJ...918L..28M}); ii)
$10.94\text{ km}<R(1.44M_{\odot})<15.50\text{ km}$ \cite{2019ApJ...887L..24M} and $10.57\text{ km}<R(1.34M_{\odot})<14.86\text{ km}$ \cite{2019ApJ...887L..21R} for PSR J0030+0451; and iii) $\tilde{\Lambda}<720$ for the GW170817 event \cite{Abbott:2018wiz}. }
\label{tab:2}
\begin{tabular}{c ccccc c ccccc }
\hline \hline 
&  \multicolumn{5}{c}{Set $dM/dR<0$} & &  \multicolumn{5}{c}{Set $dM/dR\nless0$} \\
\cline{2-6} \cline{8-12} 
&\multicolumn{2}{c}{Without constraints}  & &\multicolumn{2}{c}{With constraints} & &\multicolumn{2}{c}{Without constraints}  & &\multicolumn{2}{c}{With constraints}  \\
\cline{2-3} \cline{5-6} \cline{8-9} \cline{11-12} 
& min & max & & min & max & & min & max & & min & max \\ 
 \hline
 $M_{\mathrm{max}}$ [$M_{\odot}$] & 2.0 & 2.3 & & 2.0 & 2.3 & & 2.0 & 3.07 & & 2.0 & 2.69  \\
$R_{\mathrm{max}}$ [km] & 9.66 & 12.12 & & 10.08 & 12.12 & & 9.54 & 15.00 & & 10.07 & 13.38  \\
$R(1.4M_{\odot})$ [km] & 10.49 & 12.88 & & 11.05 & 12.88 & & 10.06 & 13.98 & & 10.92 & 13.00  \\
$n_{\mathrm{max}}$ [fm$^{-3}$] & 0.645 & 1.26 & & 0.645 & 1.212 & & 0.313 & 1.254 & & 0.424 & 1.239  \\
\hline
\hline
\end{tabular}
\end{table*}

\bibliographystyle{apsrev4-1}
%\bibliography{biblio}
\providecommand{\noopsort}[1]{}\providecommand{\singleletter}[1]{#1}%

\end{document}